\documentclass[preprint,12pt]{elsarticle}

\usepackage{amssymb}

\usepackage{amsmath,amsthm,amssymb,mathrsfs}
\usepackage{bm}

\journal{Journal of Magnetism and Magnetic Materials}

\begin{document}

\begin{frontmatter}




\title{Non-periodic input-driven magnetization dynamics in voltage-controlled parametric oscillator}


\author{Tomohiro Taniguchi 
}


\address{
 National Institute of Advanced Industrial Science and Technology (AIST), Research Center for Emerging Computing Technologies, Tsukuba, Ibaraki 305-8568, Japan, 
}



\begin{abstract}
Input-driven dynamical systems have attracted attention because their dynamics can be used as resources for brain-inspired computing. 
The recent achievement of human-voice recognition by spintronic oscillator also utilizes an input-driven magnetization dynamics. 
Here, we investigate an excitation of input-driven chaos in magnetization dynamics by voltage controlled magnetic anisotropy effect. 
The study focuses on the parametric magnetization oscillation induced by a microwave voltage and investigates the effect of random-pulse input on the oscillation behavior. 
Solving the Landau-Lifshitz-Gilbert equation, temporal dynamics of the magnetization and its statistical character are evaluated. 
In a weak perturbation limit, the temporal dynamics of the magnetization are mainly determined by the input signal, which is classified as input-driven synchronization. 
In a large perturbation limit, on the other hand, chaotic dynamics are observed, where the dynamical response is sensitive to the initial state. 
The existence of chaos is also identified by the evaluation of the Lyapunov exponent. 
\end{abstract}


\end{frontmatter}





\section{Introduction}
\label{sec:Introduction}

After the successful reports on human-voice recognition by spin-torque oscillator \cite{torrejon17}, associative memory operation by three-terminal magnetic memory \cite{borders17}, and pattern recognition by an array of spin-Hall oscillators \cite{kudo17} in 2017, the application of spintronics technology to emerging computing has become an exciting topic in magnetism \cite{grollier20,nakajima21}. 
The works bridge the research field to the others such as computer science, statistical physics, and nonlinear science. 
Among them, the input-driven dynamical theory \cite{manjunath12} has gained great attention because most models related to emerging computing, such as machine learning and robotics, are input-driven. 
For example, the human-voice recognition task can be solved using spin-torque oscillator \cite{torrejon17} if there is one-to-one correspondence between the input electric voltage, converted from human voice, and the output power originated from nonlinear magnetization dynamics. 
The correspondence as such is classified as input-driven synchronization  \cite{mainen95,toral01,teramae04,goldobin05,nakao07,imai22}, where the dynamical output from the oscillator is solely determined by the input data and is independent of the initial state of the magnetization; therefore, by learning the correspondence, the system can recognize the input data. 
Another example of the input-driven dynamics is chaos, which has a sensitivity to the initial state and has been found in brain activities and artificial neural networks \cite{aihara90,strogatz01}. 
Contrary to the input-driven synchronization in magnetization dynamics \cite{torrejon17,tsunegi18,tsunegi19,riou19,yamaguchi20,yamaguchi20srep,akashi20}, however, the input-driven chaotic dynamics in spintronics devices have not been fully investigated yet \cite{akashi20}. 


The input-driven magnetization synchronization has been mainly studied in spin-torque oscillator \cite{torrejon17,tsunegi18,tsunegi19,riou19,yamaguchi20,yamaguchi20srep,akashi20}, where electric current drives the dynamics. 
From viewpoint of energy-saving computing, it would be preferable to drive magnetization dynamics by voltage controlled magnetic anisotropy (VCMA) effect \cite{weisheit07,duan08,maruyama09,nakamura09,tsujikawa09,shiota09,nozaki10,endo10,miwa17}. 
The VCMA effect arises from the modification of electron states \cite{nakamura09,tsujikawa09} and/or the induction of magnetic moment \cite{miwa17} near the ferromagnetic/insulator interface by an application of electric voltage, and is expected to provide low-power writing scheme in magnetoresistive random access memory. 
A recognition task of the random input signal by using the relaxation dynamics of the magnetization caused by VCMA effect was reported recently \cite{taniguchi22}. 
Remind that recognition tasks are solved in terms of input-driven synchronization. 
In such circumstances, it is of interest to investigate a possibility to induce the input-driven chaos in magnetization dynamics manipulated by VCMA effect.


In this work, we propose a method to excite the input-driven chaotic magnetization dynamics in a parametric oscillator maintained by a microwave VCMA effect. 
Note that the relaxation dynamics of the magnetization caused by a direct VCMA effect may not be suitable for inducing chaos because the dynamics saturates to a fixed point, while chaos, on the other hand, must be sustained. 
To overcome the issue, we focus on the parametric magnetization oscillation caused by a microwave VCMA effect, which was recently demonstrated experimentally \cite{yamamoto20,imamura20}. 
Specifically, we study the modulation of the parametric oscillation caused by the injection of input signal and solving the Landau-Lifshitz-Gilbert (LLG) equation. 
It is shown that the magnetization dynamics in the presence of random input signal become sensitive to the initial state, indicating the appearance of input-driven chaos. 
The appearance of chaos is also investigated by evaluating the Lyapunov exponent.



\begin{figure}
\centerline{\includegraphics[width=1.0\columnwidth]{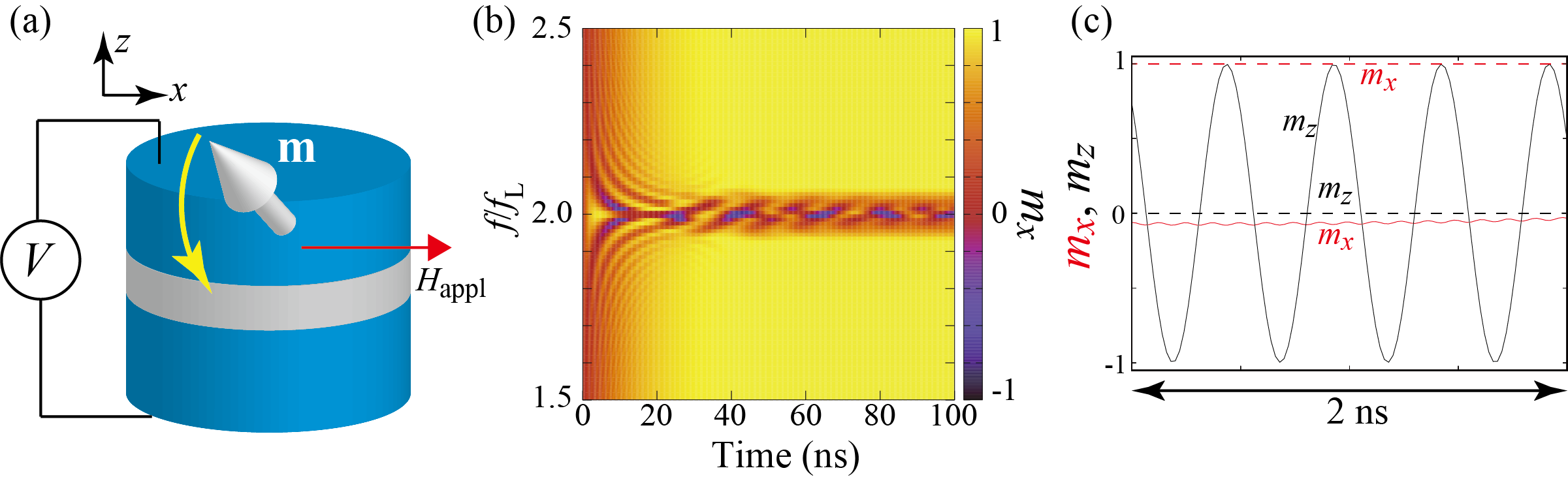}}
\caption{
            (a) Schematic illustration of a magnetic multilayer.
                 The unit vector pointing in the magnetization direction in the free layer is denoted as $\mathbf{m}$. 
                 An external magnetic field $H_{\rm appl}$ is applied in the $x$ direction. 
                 In parametric oscillation state, the magnetization rotates around the $x$ axis, as schematically shown by the yellow arrow. 
            (b) Time evolution of $m_{x}$ in the presence of a microwave voltage. 
                 The horizontal axis represents the ratio of the frequency $f$ of the voltage with respect to the Larmor frequency $f_{\rm L}$.
            (c) Examples of $m_{x}$ (red) and $m_{z}$ (black) in steady states. 
                 The solid and dotted lines correspond to the microwave frequency of $f=2.0 f_{\rm L}$ and $f=2.5 f_{\rm L}$, respectively.  
         \vspace{-3ex}}
\label{fig:fig1}
\end{figure}


\section{Temporal dynamics}
\label{sec:Temporal dynamics}

Here, we show the temporal dynamics of the magnetization in the presence of time-dependent inputs. 


\subsection{Parametric oscillation}
\label{sec:Parametric oscillation}

Figure \ref{fig:fig1}(a) shows a schematic view of a ferromagnetic multilayer consisting of free and reference layers separated by a thin nonmagnetic spacer. 
The unit vector pointing in the magnetization direction in the free layer is denoted as $\mathbf{m}$. 
The $z$ axis is normal to the film plane. 
It was experimentally confirmed \cite{yamamoto20} that the magnetization dynamics driven by VCMA effect is well described by the macrospin LLG equation, 
\begin{equation}
  \frac{d \mathbf{m}}{dt}
  =
  -\gamma
  \mathbf{m}
  \times
  \mathbf{H}
  +
  \alpha
  \mathbf{m}
  \times
  \frac{d \mathbf{m}}{dt}, 
  \label{eq:LLG}
\end{equation}
where $\gamma$ and $\alpha$ are the gyromagnetic ratio and the Gilbert damping constant, respectively. 
The magnetic field $\mathbf{H}$ consists of the in-plane external magnetic field $H_{\rm appl}$ and the perpendicular magnetic anisotropy field $H_{\rm K}$ as \cite{yamamoto20} 
\begin{equation}
  \mathbf{H}
  =
  H_{\rm appl}
  \mathbf{e}_{x}
  +
  H_{\rm K}
  m_{z}
  \mathbf{e}_{z}, 
  \label{eq:field}
\end{equation}
where $\mathbf{e}_{i}$ ($i=x,y,z$) is the unit vector in the $i$-direction and we assume that the external magnetic field points to the $x$ direction. 
The values of the parameters are similar to those used in Refs. \cite{yamamoto20,imamura20}, where $\gamma=1.764\times 10^{7}$ rad/(Oe s), $\alpha=0.005$, and $H_{\rm appl}=720$ Oe. 
Note that, when $H_{\rm appl}$ and $H_{\rm K}$ are constants, the magnetization dynamics described by Eq. (\ref{eq:LLG}) are relaxation dynamics towards the minima of the energy density $E=-M \int d\mathbf{m}\cdot\mathbf{H}$, i.e., the magnetization saturates to a fixed point. 
Therefore, to excite sustainable dynamics such as an oscillation or chaos, $H_{\rm appl}$ and/or $H_{\rm K}$ should be time-dependent. 


Let us first show the parametric oscillation of the magnetization \cite{yamamoto20,imamura20}. 
Before applying voltage, the magnetic anisotropy field $H_{\rm K}$ has a value determined by the competition between the shape and interfacial magnetic anisotropy fields \cite{yakata09,ikeda10,kubota12}. 
Next, both direct and microwave voltages are applied, which make the magnetic anisotropy field as $H_{\rm K}=H_{\rm Kd}+H_{\rm Ka}\sin(2\pi ft)$ by VCMA effect, where $H_{\rm Ka}$ and $f$ are the amplitude and frequency of the microwave component in VCMA fields. 
For simplicity, we assume that the direct component $H_{\rm Kd}$ in $H_{\rm K}$ in the presence of VCMA effect is zero \cite{imamura20}, while $H_{\rm Ka}=100$ Oe. 
Note that the value of $H_{\rm Ka}/H_{\rm appl}$ should be larger than $2\alpha$ to excite a sustainable oscillation \cite{imamura20}.
Figure \ref{fig:fig1}(b) shows the time evolution of $m_{x}$ for various $f$. 
The magnetization basically saturates to a fixed point $m_{x}=+1$ due to the relaxation to the direction of the external magnetic field. 
An exception occurs when the input frequency $f$ is close to $2 f_{\rm L}$, where $f_{\rm L}=\gamma H_{\rm appl}/(2\pi)$ is the Larmor precession frequency. 
Initially, $m_{x}$ oscillates around $m_{x}=0$ and finally tends to $m_{x}\simeq 0$. 
Figure \ref{fig:fig1}(c) summarizes the time evolution of $m_{x}$ (red) and $m_{z}$ (black) for $f=2.0 f_{\rm L}$ (solid) and $2.5 f_{\rm L}$ (dotted). 
A steady precession is excited for $f=2f_{\rm L}$, where the magnetization oscillates almost in the $yz$ plane ($m_{x}\simeq 0$); see also \ref{sec:AppendixA} showing the spatial trajectory of the oscillation. 
Since the input frequency is two times larger than the Larmor precession frequency, the oscillation is classified to the parametric oscillation. 


\subsection{Input-driven dynamics}
\label{sec:Input-driven dynamics}

Next, let us consider the input-driven dynamics. 
The microwave voltage inducing the parametric oscillation is input signal of one kind. 
In fact, it causes a synchronized motion of the magnetization with respect to the microwave voltage, where the relative phase between them saturates to one of two stable values \cite{imamura20}; see also \ref{sec:AppendixA}. 
Multistability and chaotic behavior were also found very recently \cite{celada22}. 
Such a periodic input-driven dynamics has been studied for a long time \cite{pikovsky03}. 
Note, however, that the input signal used in emerging computing is often non-periodic, as in the case of human voice. 
A main focus in recent input-driven dynamical theory \cite{manjunath12} is to study whether the dynamical response caused by non-periodic input is solely determined by the input signal or depends on the initial state of the physical system. 
The former is the input-driven synchronization. 
In the latter case, the dynamics might be the case of the input-driven chaos. 


\begin{figure}
\centerline{\includegraphics[width=1.0\columnwidth]{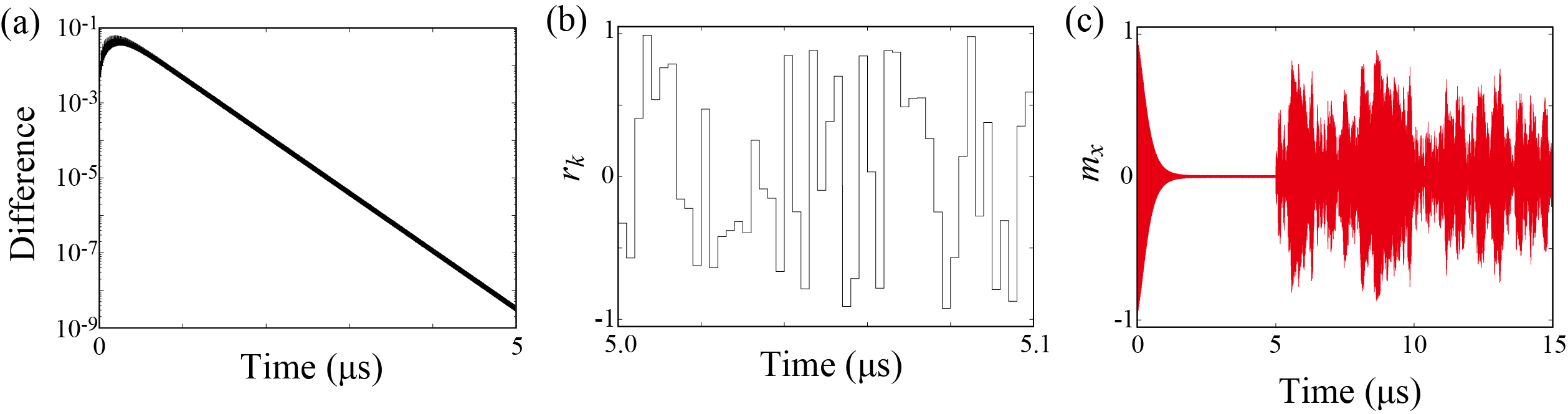}}
\caption{ 
            (a) Time evolution of the difference of two solutions of Eq. (\ref{eq:LLG}) with different initial conditions. 
            (b) Examples of the uniformly distributed random input signal $r_{k}$. 
            (c) Time evolution of $m_{x}$. 
                 The random input signal is injected from $t=5.0$ $\mu$s. 
                 The strength and the pulse width of the random input signal are $\nu=0.8$ and $t_{\rm p}=2.0$ ns, respectively. 
         \vspace{-3ex}}
\label{fig:fig2}
\end{figure}


Therefore, there are two requirements for studying the input-driven dynamics.  
First, it is necessary to compare the solutions of Eq. (\ref{eq:LLG}) with different initial conditions. 
Second, non-periodic input signal should be added to VCMA effect. 
For the first requirement, we prepare natural initial conditions in the absence of VCMA effect from thermal equilibrium distribution \cite{imai22}; see \ref{sec:AppendixB}.  
We solve the LLG equations for these initial conditions with $H_{\rm K}=H_{\rm Ka}\sin(2\pi f_{\rm L}t)$ from $t=0$ to $t=5.0$ $\mu$s, where non-periodic input is not injected yet. 
For convenience, let us denote two solutions of Eq. (\ref{eq:LLG}) with slightly different initial conditions as $\mathbf{m}_{1}$ and $\mathbf{m}_{2}$. 
Figure \ref{fig:fig2}(a) shows the time evolution of their difference, $|\mathbf{m}_{1}-\mathbf{m}_{2}|=\sqrt{(m_{1x}-m_{2x})^{2}+(m_{1y}-m_{2y})^{2}+(m_{1z}-m_{2z})^{2}}$, in the presence of a microwave voltage. 
The difference decreases with time increasing because the microwave voltage tends to fix the phase of the magnetization oscillation \cite{imamura20}.  
Simultaneously, we should note that a tiny difference still remains because the phase fixing by the microwave voltage is achieved only in the limit of $t\to\infty$. 
Next, for the second requirement, we add uniformly distributed random-pulse number $r_{k}$ $(-1 \le r_{k} \le 1)$ as input signal, which is used in a recognition task of physical reservoir computing \cite{nakajima21,akashi20}. 
The suffix $k$ represents the order of the input signal. 
Thus, from $t=5.0$ $\mu$s, the magnetic anisotropy field becomes 
\begin{equation}
  H_{\rm K}
  =
  H_{\rm Ka}
  \left(
    1
    +
    \nu r_{k}
  \right)
  \sin(2\pi f_{\rm L}t), 
  \label{eq:HK}
\end{equation}
where the frequency $f$ is fixed to $2 f_{\rm L}$. 
The dimensionless parameter $\nu$ determines the modulation of VCMA effect by the input signal. 
Figure \ref{fig:fig2}(b) shows an example of the random input signal $r_{k}$, where the pulse width is $2.0$ ns. 
The input signal modulates the magnetic anisotropy field and induces complex dynamics of the magnetization, as shown in Fig. \ref{fig:fig2}(c), where $\nu$ is $0.8$. 


\begin{figure}
\centerline{\includegraphics[width=1.0\columnwidth]{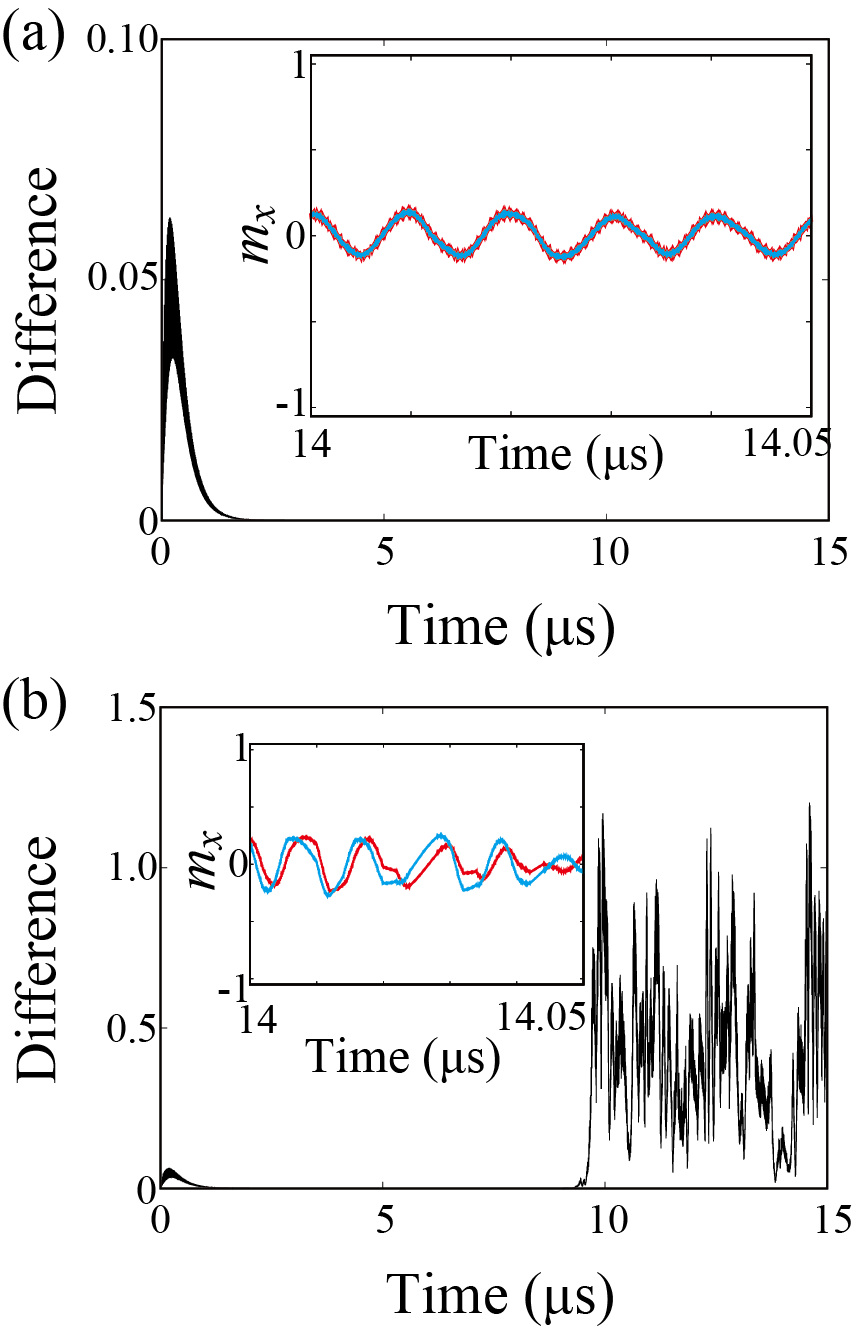}}
\caption{
            Difference $|\mathbf{m}_{1}-\mathbf{m}_{2}|$ of the solutions of the LLG equation with slightly different initial conditions for (a) $\nu=0.2$ and (b) $\nu=0.8$. 
            The insets show temporal dynamics of $m_{1x}$ and $m_{2x}$. 
         \vspace{-3ex}}
\label{fig:fig3}
\end{figure}


Now let us investigate the sensitivity of the magnetization dynamics with respect to the initial state. 
Figure \ref{fig:fig3}(a) shows the temporal difference between two solutions of Eq. (\ref{eq:LLG}) with different initial conditions, where $\nu=0.2$. 
As mentioned, for $t\le 5.0$ $\mu$s, only the microwave voltage is applied, and the difference tends to be zero. 
There is, however, still a tiny difference, as shown in Fig. \ref{fig:fig2}(a). 
This difference can be regarded as the difference given to the initial state for the dynamics in the presence of the random input signal. 
Note that, even after the injection of the random input signal from $t=5.0$ $\mu$s, the difference remains negligible for this weak ($\nu=0.2$) perturbation limit. 
The result indicates that the synchronization caused by the microwave VCMA effect is maintained. 
The conclusion can be verified from a different viewpoint shown in the inset of Fig. \ref{fig:fig3}(a), where two solutions of the LLG equation are almost overlapped. 
However, when the strength of the random input signal becomes large as $\nu=0.8$, a tiny difference at $t=5$ $\mu$s is enlarged due to the excitation by the random input, as shown in Fig. \ref{fig:fig3}(b). 
Remind that the LLG equation conserves the norm of the magnetization as $|\mathbf{m}|=1$; therefore, the maximum value of the difference between two solutions is $2$, at which two magnetizations point to the opposite direction. 
Therefore, the difference shown in Fig. \ref{fig:fig3}(b), which is larger than $1$, is regarded as non-negligible. 
The temporal dynamics of two solutions shown in the inset of the figure also indicate that the synchronization caused by the microwave VCMA effect is broken. 
These results indicate that, although the difference of two solutions at $t=5$ $\mu$s is negligibly small, as shown in Fig. \ref{fig:fig2}(a), it is expanded by the injection of the random-pulse input signal. 
In other words, the dynamics are sensitive to the difference at $t=5$ $\mu$s. 
Such a sensitivity implies that the dynamics in Fig. \ref{fig:fig3}(b) is chaos.


\subsection{Validity of parameters}
\label{sec:Validity of parameters}

We note that the value of the parameters used in this work is in a reasonable range realized in experiments. 
The perpendicular magnetic anisotropy energy density,$K$, consists of the bulk magnetic anisotropy energy density $K_{V}$, the interfacial magnetic anisotropy energy $K_{\rm i}$, the contribution from the VCMA effect as $Kd=K_{V}d+K_{\rm i}-\eta\mathscr{E}$, where $d$ is the thickness of the free layer. 
The electric field $\mathscr{E}$ relates to the voltage $V$ via $\mathscr{E}=V/d_{\rm I}$, where $d_{\rm I}$ is the thickness of the insulating barrier. 
In typical magnetic multilayers, where the free layer and insulating barrier are CoFeB and MgO, respectively, $K_{\rm i}$ is the dominant contribution to $K$ and its value increases with the composition of Fe increasing \cite{yakata09}. 
It can reach on the order of $1.0$ mJ/m${}^{2}$, which corresponds to, typically, on the order of $1$ T in terms of magnetic field, $2K_{\rm i}/(Md)$, where $M$ is the saturation magnetization and is about $1000$ emu/cm${}^{3}$. 
On the other hand, the VCMA efficiency $\eta$ reaches $300$ fJ/(Vm) \cite{nozaki17,nozaki20}. 
Regarding typical values of the thickness of the insulating barrier (about $2.5$ nm) and applied voltage ($0.5$ V at maximum) \cite{sugihara19}, the tunable range of the magnetic anisotropy field by voltage is on the order of $1.0$ kOe at maximum. 
Summing these values, it is possible to generate an oscillating component of the magnetic anisotropy field on the order of $100$ Oe. 
It should also be noted that a series of random-pulse input signal with the pulse width of nanoseconds was applied to magnetic multilayers in experiments of physical reservoir computing \cite{tsunegi18,tsunegi19}. 
Therefore, the proposal made here will be experimentally examined. 


\subsection{Comment on LLB equation}
\label{sec:Comment on LLB equation}

The results shown in this work are derived by solving the LLG equation. 
There is another equation of motion, Landau-Lifshitz-Bloch (LLB) equation, describing the magnetization dynamics. 
Here, let me mention their differences briefly. 

The LLG equation assumes the conservation of the magnetization magnitude, i.e., $|\mathbf{m}|=1$, which is valid at temperature sufficiently lower than Curie temperature. 
The relaxation of the magnetization is characterized by the dimensionless damping parameter $\alpha$. 
Note that the number of independent variables in the LLG equation is two, although the vector $\mathbf{m}$ has three components in the Cartesian coordinate. 
This is because the condition $|\mathbf{m}|=1$ acts as a constraint and reduces the number of independent variables. 
On the other hand, the LLB equation does not conserve the magnetization magnitude, and is valid at high temperature. 
There are two parameters, the longitudinal and transverse relaxation times, characterizing the magnetization relation. 
The number of independent variables is three in the LLB equation. 

We should note that chaos appears in a high-dimensional system. 
In fact, chaos is prohibited in a dynamical system whose dimension is less or equal to two, according to the Poincar\'e-Bendixson theorem. 
Therefore, chaos might be easily excited in a system described by the LLB equation than that described by the LLG equation. 
However, since the number of the parameters describing the relaxation are different between two equations, it is difficult to compare chaos in these two equations on an equal footing. 
Therefore, we would like to leave chaos in the LLB equation for further study in future.


\section{Statistical analysis of Lyapunov exponent}
\label{sec:Statistical analysis of Lyapunov exponent}

In Sec. \ref{sec:Parametric oscillation}, we study the existence of chaos from temporal dynamics. 
To identify chaos from different perspectives, here, we evaluate the Lyapunov exponent.


\begin{figure}
\centerline{\includegraphics[width=1.0\columnwidth]{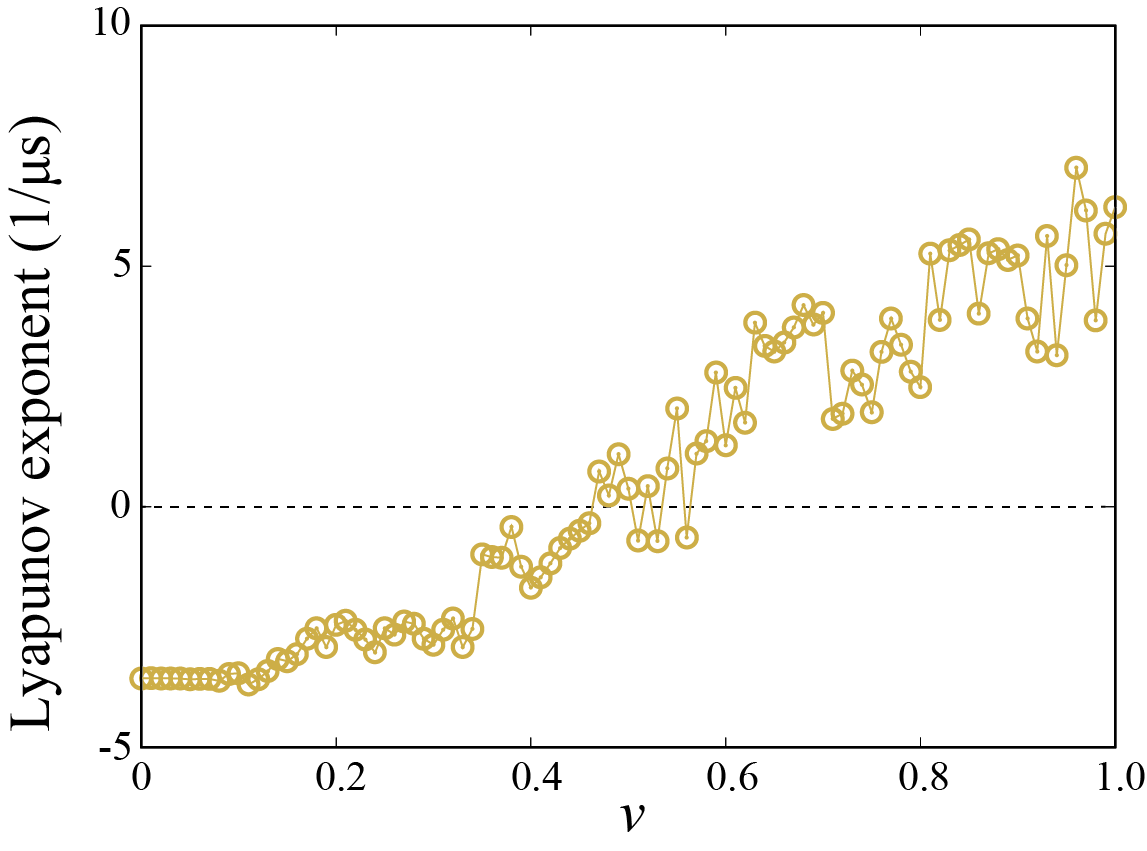}}
\caption{
             Lyapunov exponent as a function of the dimensionless input strength $\nu$. 
         \vspace{-3ex}}
\label{fig:fig4}
\end{figure}


The Lyapunov exponent is an expansion rate of the difference between two solutions of an equation of motion with slightly different initial conditions. 
The Lyapunov exponent is negative when the solution saturates to a fixed point. 
The input-driven synchronization is an example of the dynamics with a negative Lyapunov exponent because the temporal dynamics are solely determined by the input signal and independent of the initial condition. 
When the Lyapunov exponent is zero, the difference remains constant. 
An example of the dynamics corresponding to a zero Lyapunov exponent is a limit-cycle oscillation. 
The corresponding dynamics thus depends on the initial state but is not chaos. 
When the Lyapunov exponent is positive, the difference is expanded and thus, the dynamics are sensitive to the initial state. 
A positive Lyapunov exponent indicates an existence of chaos. 
Note that the sensitivity to the initial state in dynamics is a necessary condition of chaos but is not a sufficient condition because the dynamics with a zero Lyapunov exponent also depends on the initial state. 
The evaluation of the Lyapunov becomes a measure of chaos because its sign provides an evidence of chaos. 
Here, we evaluate the Lyapunov exponent by Shimada-Nagashima method \cite{shimada79}, where the exponent is defined as 
\begin{equation}
  \varLambda
  =
  \lim_{N \to \infty}
  \frac{1}{N \Delta t}
  \sum_{i=1}^{N}
  \ln
  \frac{\mathscr{D}}{\epsilon},
  \label{eq:Lyapunov_exponent}
\end{equation}
where $\Delta t$ is the time increment of the LLG equation and is $1$ ps in this work. 
In the Shimada-Nagashima method, the solution of an equation of motion at a certain time $t_{0}$ is shifted with a tiny distance $\epsilon$ in phase space. 
Then, the original and shifted solutions are evolved from $t=t_{0}$ to $t=t_{0}+\Delta t$ by the equation of motion. 
The distance between these solutions at $t=t_{0}+\Delta t$ is $\mathscr{D}$. 
If $\mathscr{D}/\epsilon<(>)1$, the difference given at the time $t_{0}$ shrinks (expanded), and thus, the temporal Lyapunov exponent is negative (positive). 
The Lyapunov exponent is a long-time average of such a temporal Lyapunov exponent, as implied by Eq. (\ref{eq:Lyapunov_exponent}); see also \ref{sec:AppendixC} for details. 
Figure \ref{fig:fig4} summarizes the Lyapunov exponent as a function of the strength of the input signal, $\nu$. 
For small $\nu$, the Lyapunov exponent is negative, indicating that the dynamical state of the magnetization is determined by the input signal and is insensitive to the initial sate. 
The Lyapunov exponent changes its sign around $\nu=0.5$ and becomes positive for large $\nu$, indicating that the dynamics becomes sensitive to the initial state. 
The positive Lyapunov exponents are another evidence of the appearance of input-driven chaos in the parametric oscillator.




\section{Conclusion}
\label{sec:Conclusion}

In conclusion, the input-driven magnetization dynamics in the parametric oscillator were studied by solving the LLG equation. 
The microwave voltage induces a sustainable oscillation of the magnetization around an external magnetic field through VCMA effect. 
Adding non-periodic input signal changes the dynamical behavior, depending on its magnitude. 
In a weak perturbation limit, the temporal dynamics of the magnetization were determined by the input signal and are insensitive to the initial state. 
On the other hand, in a large perturbation limit, the dynamics become sensitive to the initial state. 
Such a chaotic behavior was revealed by comparing the difference of two solutions of the LLG equation with different initial conditions. 
The evaluation of the Lyapunov exponent also identified the appearance of chaos in the magnetization dynamics. 

The existence of chaos in the input-driven spintronics systems will be of interest for emerging computing technologies. 
For example, it has been empirically shown that the computing performance of physical reservoir computing is maximized at the edge of chaos \cite{bertschinger04,nakayama16}, although it does not seem a general conclusion \cite{nakajima21}. 
Therefore, a tunability of the dynamical state in physical systems is required for an enhancement of the computing capability. 
The result shown in Fig. \ref{fig:fig4} shows, for example, that the dynamical state of spintronics devices can be tuned between input-driven synchronization and chaos by tuning the input strength. 
As emphasized in Sec. \ref{sec:Validity of parameters}, the values of the parameters used in this work are in a reasonable range available in experiments, and therefore, the results in this work will provide a direction to design the emerging computing devices based on spintronics technologies. 
The input-driven chaotic magnetization dynamics might also have some applications because chaos was found in brain activities \cite{strogatz01} and theoretical models emulating the neural dynamics of squid \cite{aihara90}. 
Developing the present results to brain-inspired computing will be, therefore, an interesting future work. 


\section*{Acknowledgments}
The work is supported by JSPS KAKENHI Grant Number 20H05655. 



  \bibliographystyle{elsarticle-num} 


\appendix


\section{Parametric oscillation by microwave voltage}
\label{sec:AppendixA}

\begin{figure}
\centerline{\includegraphics[width=1.0\columnwidth]{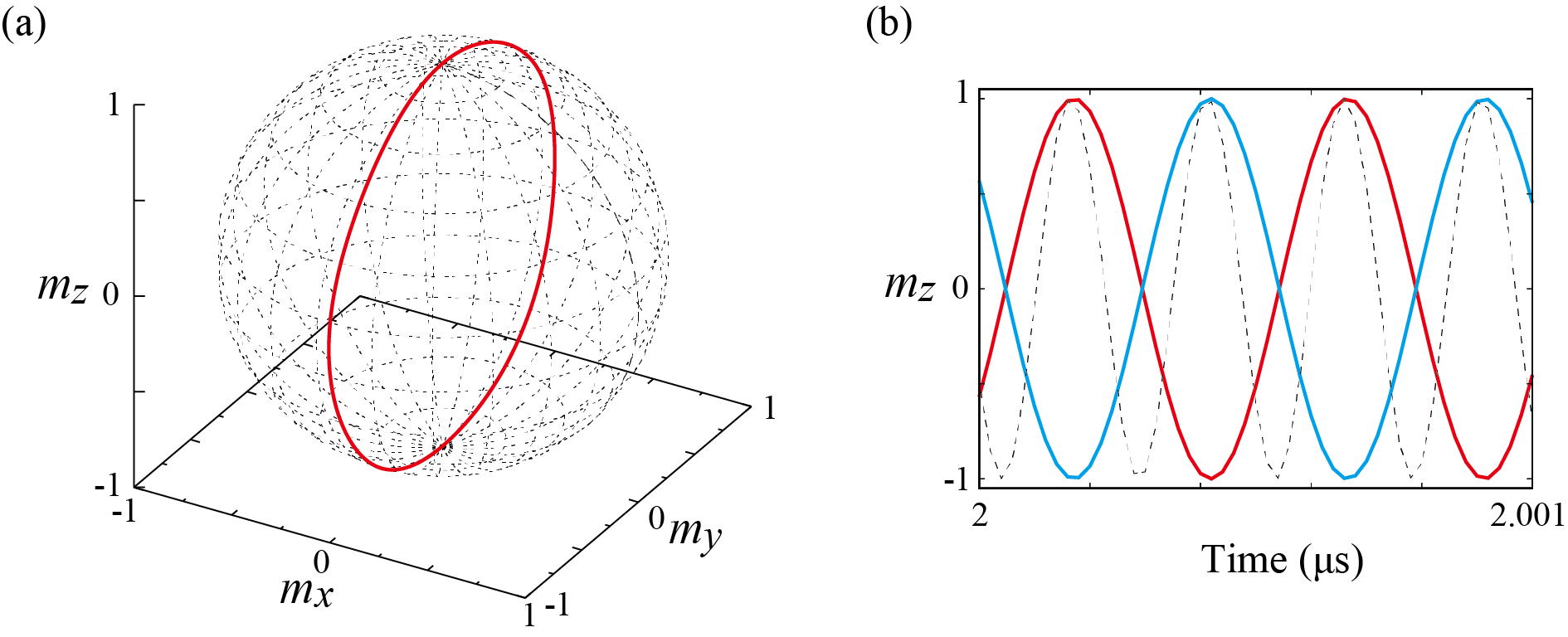}}
\caption{
            (a) Spatial trajectory of the parametric oscillation induced by a microwave voltage. 
            (b) Temporal evolution of $m_{z}$ with different initial conditions. 
                 Dotted line represents the oscillation of the microwave voltage for comparison. 
         \vspace{-3ex}}
\label{fig:fig5}
\end{figure}

In the main text, two time-dependent inputs are added to the magnetic anisotropy field. 
One is a microwave voltage and the other is uniformly distributed random numbers. 
The former induces a parametric oscillation \cite{yamamoto20}. 
Figure \ref{fig:fig5}(a) shows the spatial trajectory of the magnetization oscillation in a steady state. 
As mentioned in the main text, the magnetization oscillates around the $x$ axis. 
The solid lines in Fig. \ref{fig:fig5}(b) show examples of the magnetization oscillation with different initial conditions, whereas the dotted line represents the oscillation of the microwave voltage, $\sin(2\pi f_{\rm L}t)$. 
It indicates that the oscillation frequency is a half of the microwave frequency. 


The microwave voltage fixes the phase of the magnetization with respect to the voltage oscillation. 
There are more than one solution of the magnetization phase \cite{imamura20}. 
The phase depends on the initial conditions, as implied in Fig. \ref{fig:fig5}(b); see also Sec. \ref{sec:AppendixB} below. 
When we study chaos in the main text, we choose the solutions of the LLG equation with the same phases because chaos is characterized by the sensitivity to the initial state.


\section{Preparation of initial state}
\label{sec:AppendixB}

Chaotic dynamics are sensitive to the initial state. 
Therefore, to identify the existence of chaos, it is necessary to study the dependence of the temporal dynamics on the initial state. 
We prepare natural initial states by solving the LLG equation in the absence of the input signal. 
The value of $H_{\rm K}$ is that in the absence of external voltage and is $6.28$ kOe \cite{yamamoto20}. 
Also note that, at zero temperature, the solution of the LLG equation saturates to the minimum energy state, $\sin\theta=H_{\rm appl}/H_{\rm K}$, where $\theta$ relates to $m_{z}$ via $m_{z}=\cos\theta$. 
To obtain natural distribution of the initial state \cite{imai22}, we add a torque, $-\gamma\mathbf{m}\times\mathbf{h}$, due to thermal fluctuation to the right-hand side of Eq. (\ref{eq:LLG}). 
Here, the components of $\mathbf{h}$ satisfy the fluctuation-dissipation theorem \cite{brown63}, 
\begin{equation}
  \langle h_{k}(t) h_{\ell}(t^{\prime}) \rangle 
  =
  \frac{2\alpha k_{\rm B}T}{\gamma MV}
  \delta_{k\ell}
  \delta(t-t^{\prime}), 
\end{equation}
where the saturation magnetization $M$ is assumed to be $955$ emu/cm${}^{2}$ \cite{yamamoto20}. 
The temperature $T$ is $300$ K, while the volume is $V=\pi \times 50 \times 50 \times 1.1$ nm${}^{3}$, which is typical for VCMA experiments. 
The thermal fluctuation excites a small-amplitude oscillation of the magnetization around the energetically minimum state with the ferromagnetic resonance frequency. 
We pick up the temporal directions of the oscillating magnetization and use them as the natural initial states.


\begin{figure}
\centerline{\includegraphics[width=1.0\columnwidth]{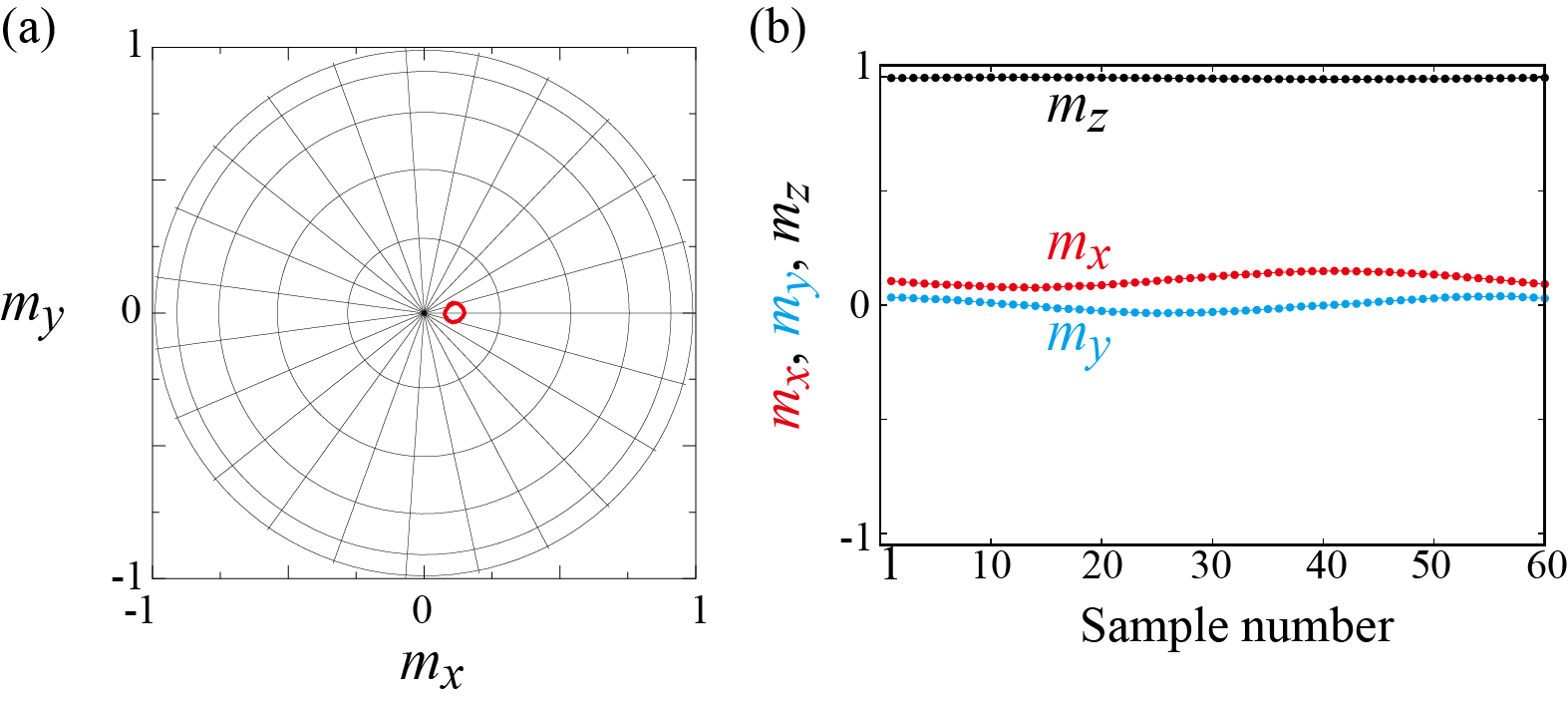}}
\caption{
            (a) Spatial distribution of the initial states prepared by solving the LLG equation with thermal fluctuation. 
            (b) The samples of $m_{x}$, $m_{y}$, and $m_{z}$ corresponding to the small-amplitude oscillation of the magnetization around the energetically minimum state excited by thermal fluctuation.  
         \vspace{-3ex}}
\label{fig:fig6}
\end{figure}


Figure \ref{fig:fig6}(a) shows the spatial distribution of the initial states, where we prepared $60$ samples. 
Figure \ref{fig:fig6}(b) summarize the values of $\mathbf{m}$ for these samples. 
For example, the dynamics shown in Fig. 3 in the main text are derived by using the sample numbers $1$ and $2$ as the initial states, where the solutions of the magnetization in both samples have the same phase when the dynamics are driven by a microwave voltage. 
On the other hand, in Fig. \ref{fig:fig5}(b), the red and blue lines correspond to the sample number $1$ and $15$. 
They are unsuitable to study chaos because the dynamical states at which the random input signal is injected are greatly different. 


\section{Evaluation method of Lyapunov exponent}
\label{sec:AppendixC}

The Lyapunov exponent is evaluated by Shimada-Nagashima method \cite{shimada79}. 
As written in the main text, we add the random input signal from $t=5$ $\mu$s. 
Let us denote the solution of the LLG equation at this time as $\mathbf{m}(t)$. 
We introduce the zenith and azimuth angles, $\theta$ and $\varphi$, as $\mathbf{m}=(m_{x},m_{y},m_{z})=(\sin\theta\cos\varphi,\sin\theta\sin\varphi,\cos\theta)$, i.e., $\varphi=\tan^{-1}(m_{y}/m_{x})$ and $\theta=\cos^{-1}m_{z}$. 
Then, we also introduce $\mathbf{m}^{(1)}(t)=(\sin\theta^{(1)}\cos\varphi^{(1)},\sin\theta^{(1)}\sin\varphi^{(1)},\cos\theta^{(1)})$.
Here, $\theta^{(1)}$ and $\varphi^{(1)}$ satisfy $\epsilon=\sqrt{[\theta-\theta^{(1)}]^{2}+[\varphi-\varphi^{(1)}]^{2}}$, where $\epsilon=1.0\times 10^{-5}$ is a fixed value. 
For convenience, let us introduce a notation, 
\begin{equation}
  \mathcal{D}[\mathbf{m}(t),\mathbf{m}^{(1)}(t)]
  =
  \sqrt{
    \left[
      \theta(t)
      -
      \theta^{(1)}(t)
    \right]^{2}
    +
    \left[
      \varphi(t)
      -
      \varphi^{(1)}(t)
    \right]^{2}
  }
\end{equation}
Solving the LLG equations of $\mathbf{m}(t)$ and $\mathbf{m}^{(1)}(t)$, we obtain $\mathbf{m}(t+\Delta t)$ and $\mathbf{m}^{(1)}(t+\Delta t)$. 
From them, we evaluate 
\begin{equation}
  \mathcal{D}[\mathbf{m}(t+\Delta t),\mathbf{m}^{(1)}(t+\Delta t)]
  =
  \sqrt{
    \left[
      \theta(t+\Delta t)
      -
      \theta^{(1)}(t+\Delta t)
    \right]^{2}
    +
    \left[
      \varphi(t+\Delta t)
      -
      \varphi^{(1)}(t+\Delta t)
    \right]^{2}
  }
\end{equation}
Then, a temporal Lyapunov exponent at $t+\Delta t$ is given by 
\begin{equation}
  \varLambda^{(1)}
  =
  \frac{1}{\Delta t}
  \ln
  \frac{\mathscr{D}^{(1)}}{\epsilon},
\end{equation}
where $\mathscr{D}^{(1)}=\mathcal{D}[\mathbf{m}(t+\Delta t),\mathbf{m}^{(1)}(t+\Delta t)]$. 

Next, we introduce $\mathbf{m}^{(2)}(t+\Delta t)=(\sin\theta^{(2)}\cos\varphi^{(2)},\sin\theta^{(2)}\sin\varphi^{(2)},\cos\theta^{(2)})$, where $\theta^{(2)}$ and $\varphi^{(2)}$ are defined as 
\begin{align}
&
  \theta^{(2)}(t+\Delta t)
  =
  \theta(t+\Delta t)
  +
  \epsilon
  \frac{\theta^{(1)}(t+\Delta t)-\theta(t+\Delta t)}{\mathcal{D}[\mathbf{m}(t+\Delta t),\mathbf{m}^{(1)}(t+\Delta t)]},
\\
&
  \varphi^{(2)}(t+\Delta t)
  =
  \varphi(t+\Delta t)
  +
  \epsilon
  \frac{\varphi^{(1)}(t+\Delta t)-\varphi(t+\Delta t)}{\mathcal{D}[\mathbf{m}(t+\Delta t),\mathbf{m}^{(1)}(t+\Delta t)]}.
\end{align}
According to these definitions, we notice that 
\begin{equation}
  \mathcal{D}[\mathbf{m}(t+\Delta t),\mathbf{m}^{(2)}(t+\Delta t)]
  =
  \epsilon.
\end{equation}
In other words, $\mathbf{m}^{(2)}(t+\Delta t)$ is defined by moving $\mathbf{m}(t+\Delta t)$ to the direction of $\mathbf{m}^{(1)}(t+\Delta t)$ with a distance $\epsilon$ in the $(\theta,\varphi)$ phase space. 
Then, we solve the LLG equations for $\mathbf{m}(t+\Delta t)$ and $\mathbf{m}^{(2)}(t+\Delta t)$ and obtain $\mathbf{m}(t+2\Delta t)$ and $\mathbf{m}^{(2)}(t+2\Delta t)$. 
The temporal Lyapunov exponent at $t+2\Delta t$ is 
\begin{equation}
  \varLambda^{(2)}
  =
  \frac{1}{\Delta t}
  \ln
  \frac{\mathscr{D}^{(2)}}{\epsilon},
\end{equation}
where $\mathscr{D}^{(2)}=\mathcal{D}[\mathbf{m}(t+2\Delta t),\mathbf{m}^{(1)}(t+2\Delta t)]$. 

These procedures are generalized. 
At $t+n\Delta t$, we have $\mathbf{m}(t+n\Delta t)=(\sin\theta(t+n\Delta t)\cos\varphi(t+n\Delta),\sin\theta(t+n\Delta t)\sin\varphi(t+n\Delta t),\cos\theta(t+n\Delta t))$ and $\mathbf{m}^{(n)}(t+n\Delta t)=(\sin\theta^{(n)}(t+n\Delta t)\cos\varphi^{(n)}(t+n\Delta),\sin\theta^{(n)}(t+n\Delta t)\sin\varphi^{(n)}(t+n\Delta t),\cos\theta^{(n)}(t+n\Delta t))$. 
Then, we define $\mathbf{m}^{(n+1)}(t+n\Delta t)=(\sin\theta^{(n+1)}(t+n\Delta t)\cos\varphi^{(n+1)}(t+n\Delta),\sin\theta^{(n+1)}(t+n\Delta t)\sin\varphi^{(n+1)}(t+n\Delta t),\cos\theta^{(n+1)}(t+n\Delta t))$ by moving $\mathbf{m}(t+n\Delta t)$ to the direction of $\mathbf{m}^{(n)}(t+n\Delta t)$ with a distance $\epsilon$ in the phase space as 
\begin{align}
&
  \theta^{(n+1)}(t+n\Delta t)
  =
  \theta(t+n\Delta t)
  +
  \epsilon
  \frac{\theta^{(n)}(t+n\Delta t)-\theta(t+n\Delta t)}{\mathcal{D}[\mathbf{m}(t+n\Delta t),\mathbf{m}^{(n)}(t+n\Delta t)]},
\\
&
  \varphi^{(n+1)}(t+n\Delta t)
  =
  \varphi(t+n\Delta t)
  +
  \epsilon
  \frac{\varphi^{(n)}(t+n\Delta t)-\varphi(t+n\Delta t)}{\mathcal{D}[\mathbf{m}(t+n\Delta t),\mathbf{m}^{(n)}(t+n\Delta t)]}. 
\end{align}
Note that $\mathcal{D}[\mathbf{m}(t+n\Delta t),\mathbf{m}^{(n+1)}(t+n\Delta t)] =\epsilon$. 
Then, solving the LLG equations of $\mathbf{m}(t+n\Delta t)$ and $\mathbf{m}^{(n+1)}(t+n\Delta t)$, we obtain $\mathbf{m}(t+(n+1)\Delta t)$ and $\mathbf{m}^{(n+1)}(t+(n+1)\Delta t)$. 
A temporal Lyapunov exponent at $t+(n+1)\Delta t$ is 
\begin{equation}
  \varLambda^{(n+1)}
  =
  \frac{1}{\Delta t}
  \ln
  \frac{\mathscr{D}^{(n+1)}}{\epsilon},
\end{equation}
where $\mathscr{D}^{(n+1)}=\mathcal{D}[\mathbf{m}(t+(n+1)\Delta t),\mathbf{m}^{(n+1)}(t+(n+1)\Delta t)]$. 
The Lyapunov exponent is defined as a long-time average of the temporal Lyapunov exponent as 
\begin{equation}
  \varLambda
  =
  \lim_{N \to \infty}
  \frac{1}{N}
  \sum_{i=1}^{N}
  \varLambda^{(i)}. 
\end{equation}

Note that $\mathbf{m}^{(1)}(t)$ given at the initial time can point in arbitrary directions, although it should satisfy the condition $\mathcal{D}[\mathbf{m}(t),\mathbf{m}^{(1)}(t)]=\epsilon$. 
Shimada-Nagashima method assumes that, even if the initial perturbation points to in arbitrary directions, the difference of two solutions will arrive at the mostly expanded direction by repeating the procedure. 
Since the random input signals are injected from $t=5.0$ $\mu$s to $t=15.0$ $\mu$s and the time increment is $1$ ps, we evaluate $(15.0-5.0)\mu{\rm s}/1{\rm p}s=10^{7}$ temporal Lyapunov exponents and evaluate the average. 








\end{document}